\newif\ifuseprd
\newif\ifom
\useprdfalse 
\omfalse    
\ifuseprd 
\documentclass[preprint,tightenlines,floats,prd,eqsecnum,nobibnotes,%
nofootinbib]{revtex4}
\else
\documentclass[final,12pt,letterpaper]{JHEP}
\fi 
\usepackage{amsmath,amssymb}

\newif\ifspinpm 
\spinpmtrue 

\allowdisplaybreaks[3]
\ifom
\input dnmacs
\fi 

\def\omt{{\ifom{{\dn\dnhalf :}}\else%
        {{3\!{\footnotesize$\mathbf{{\frown}\llap{\text{\tiny$\prime$}}}$}%
        {\hbox to -.7ex{\null}\llap{\raise1.3ex\hbox{\tiny{%
        \setbox255=\hbox{$\mathbf{{\smile}}$}%
        \copy255\kern-.7\wd255{\raise.5ex\hbox{$\mathbf{\cdot}$}}}}}}}}\fi}}

\newcommand\skipthis[1]{{}}

\let\oldAE\AE
\renewcommand\AE{{\ifmmode{\text{\it\oldAE}}\else{\oldAE}\fi}}

\newcommand\ct[1]{{\ifuseprd{\em{#1}},\else{\sf {#1}},\fi}}
\newcommand\bt[1]{{\em {#1}},}
\newcommand\web[1]{{\tt \hbox{{#1}}}}
\newcommand\phepth[1]{{\tt [\hepth{#1}]}}

\DeclareMathOperator{\diag}{diag}

\chardef\til=`~

\newcommand\half{{\ensuremath{\frac{1}{2}}}}
\newcommand\p{\ensuremath{\partial}}

\newcommand\abs[1]{\ensuremath{\left\lvert{#1}\right\rvert}}

\newcommand\order[1]{{\ensuremath{{\mathcal O}({#1})}}}
\newcommand\vev[1]{{\ensuremath{\left\langle{#1}\right\rangle}}}

\newcommand\apr{{\ensuremath{{\alpha'}}}}

\newcommand\lam{\lambda}
\newcommand\del{\partial}
\newcommand\OO{{\ensuremath{{\cal O}}}}

\newcommand\ha{{\half}}
\newcommand\ra{{\rightarrow}}
\newcommand\sF{{\ensuremath{{\cal F}}}}
\newcommand\papr{{2 \pi \apr}}

\providecommand\putabstract[1]{\ifuseprd\begin{abstract} {#1} \end{abstract}%
                           \else \abstract{{#1}} \fi}
\providecommand\plb[3]{{Phys.\ Lett.\ B {\bf {#1}}, {#3} ({#2})}}
\providecommand\npb[3]{{Nucl.\ Phys.\ {\bf B{#1}}, {#3} ({#2})}}
\providecommand\jhep[3]{{J.\ High Energy Phys.\ {\bf #1}, {#3} ({#2})}}

\providecommand\hepth[1]{{\tt hep-th/{#1}}}
\newenvironment{smaleq}{\ifuseprd\else\small\fi}{}

\ifuseprd
\begin{document} 
\fi 

\title{Supergravity Couplings of Noncommutative D-branes}
\ifuseprd
\author{Hong Liu}\email{liu@physics.rutgers.edu}
\author{Jeremy Michelson}\email{jeremy@physics.rutgers.edu}
\affiliation{New High Energy Theory Center \\ 
       Rutgers University \\
       126 Frelinghuysen Road \\
       Piscataway, NJ \ 08854}
\else 
\author{Hong Liu\thanks{\tt liu@physics.rutgers.edu} \ and Jeremy
Michelson\thanks{\tt jeremy@physics.rutgers.edu} \\
New High Energy Theory Center \\
Rutgers University \\
126 Frelinghuysen Road \\
Piscataway, NJ \ 08854 \ USA}
\fi 

\putabstract{
We discuss the supergravity couplings of noncommutative D-branes by 
considering the disk amplitudes with one closed string insertion. 
The result confirms a recent proposal for the general form of the 
noncommutative Yang-Mills operators coupling to the massless closed string 
modes. The construction involves smearing Yang-Mills field
variables along an open Wilson line. For multiple 
D-branes interacting with background supergravity fields,  
this prescription reduces, in the $B=0$ limit, 
to the ``symmetrized  trace'' prescription, 
and the supergravity fields are seen  to be functionals 
of the nonabelian scalar fields on the branes. 
}

\preprint{RUNHETC-2001-01\ifuseprd,~\else\\\fi {\tt hep-th/0101016}}

\ifuseprd
\maketitle
\else
\begin{document}
\fi 

\section{Introduction}

It has become clear that open Wilson lines~\cite{iikk,ambjorn} 
play a fundamental role in the dynamics of noncommutative gauge theories.
They are essential to building gauge invariant operators  
and are thus essential when coupling noncommutative Yang-Mills modes 
to closed string modes~\cite{dasrey,ghi}. They are also 
important for a proper understanding of the loop dynamics of
noncommutative gauge theories, and of the Seiberg-Witten map~\cite{sw} 
between ordinary and noncommutative Yang-Mills 
modes~\cite{liu} (see also~\cite{mehen}). 

An open Wilson line of momentum $k$ along some open contour $C$ 
is given by
\begin{equation} \label{mwilson}
W_k (C) =   {\rm Tr} \int d^d x \, W(x,C) \ast e^{i k \cdot x}
\end{equation}
where $W(x,C)$ is a Wilson line in which multiplication is given by
the $\ast$-product.
\skipthis{
\begin{eqnarray} \label{cwilson}
W(x,C) & = & P_{\ast} \exp \left( i 
\int_0^1 d \sigma \partial_{\sigma} \,
\xi^{\mu} (\sigma) \, \hat{A}_{\mu} (x + \xi (\sigma))
\right) 
\end{eqnarray}
and $P_{\ast}$ denotes path ordering with multiplication given by
the $\ast$-product.} 
$W_{k} (C)$ is gauge invariant~\cite{iikk,ambjorn}  when the momentum 
$k$ of the line, and the distance 
$\Delta x$ between its two end points, satisfy the relation
\begin{equation} \label{gawil}
 \Delta x^{\mu} = 
\theta^{\mu \nu} k_{\nu},
\end{equation}
where $\theta$ is the noncommutative parameter of the theory.
Similarly, for any local operator $\OO(x)$ which transforms adjointly under 
gauge transformations, $\OO(k)$, defined by%
\footnote{$P_\ast$ denotes path ordering in which multiplication given
by the $\ast$-product.}
\begin{equation} \label{insert}
\OO(k) =  {\rm Tr} \int d^d x \; P_{\ast} 
\left[ W(x,C) \OO (y) \right] \ast e^{i k \cdot x} ,
\end{equation}
is gauge invariant~\cite{ghi},
where $y$ is any point on the path $C$  of the Wilson line.
Thus whereas there is no gauge invariant local operator in position space, 
it is possible to write down 
the linearized 
coupling of a supergravity field $h$ to the noncommutative Yang-Mills modes 
in momentum space\cite{dasrey,ghi}
\begin{equation}
\int d^d k \; h(-k)\,  \OO (k) 
\end{equation}
with $\OO(k)$ an operator of type~\eqref{insert}. 

However, since inserting  local operators at any point of 
an open Wilson line all yields gauge invariant 
operators, it is not {\em a priori\/} clear what the precise ordering 
prescription  is for constructing such a  gauge invariant operator. 
In~\cite{liu}, by considering the factorization  of 
the one-loop open string amplitudes~\cite{lm2,zanon}, it was  
proposed that  the correct prescription is to integrate 
all external insertions over a straight Wilson line%
\footnote{Straight Wilson lines were first advocated in~\cite{ghi}.}.
For example, the Yang-Mills operator that couples to the dilaton contains   
\begin{equation} \label{giop}
 \int d^d x  \! 
 \int^{1}_{0}\! d \tau_1 d \tau_2 
 \; P_{\ast} \left[ W(x,C) F_{\mu \nu} (x + \xi(\tau_1)) F^{\mu \nu} (x + \xi(\tau_2))
 \right] \ast e^{i k \cdot x} 
\end{equation}
with $C$ parameterized by $\xi(\tau)$,  a straight line. 
\skipthis{In the second line we have used 
a short-hand notation $L_{\ast}$ to denote the integrations 
together with the path ordering prescription.}%
In~\cite{dastrivedi} it was argued based on the connection between 
noncommutative
gauge theories and the Matrix 
model (see e.g.\cite{bss,miao,aoki,ishibashi,iikk,ambjorn,seiberg}) 
that the integration procedure  provides a natural generalization to 
$U(\infty)$ of the ``symmetrized trace'' prescription of the Born-Infeld 
coupling to weak external supergravity fields (see 
e.g.~\cite{tseytlin,myers,taylor}). 

In this paper we investigate the disk amplitudes between one massless
closed string in the NSNS sector and the noncommutative gauge modes on
the brane, aiming to provide further support for the integration
prescription and to write down the precise couplings of the massless
closed string modes to a noncommutative D-brane.
We will address RR insertions in a forthcoming paper.\cite{lm3} \
We shall be
interested in the leading terms of the amplitudes in the low energy
limit in terms of open string metric $G_{\mu \nu}$ on the D-brane,
while exploring the effect of the noncommutative parameter
$\theta$. More explicitly, as in~\cite{lm2}, we take the low energy
limit to mean
\begin{equation} \label{limit}
\apr k_{i \mu} G^{\mu \nu} k_{j \nu} \ll 1 \qquad
\text{with}   \qquad
k_{i \mu} \theta^{\mu \nu} k_{j \nu} \sim O(1) \ .
\end{equation}
Since the closed string metric $g_{\mu \nu}$  scales differently 
with respect to $\apr$ from the open string metric, in the above 
limit~\eqref{limit},  low energy processes  in 
terms of the open string metric on the brane do not correspond to
low energy processes in the bulk that involves the closed string metric.
This differs from the $B=0$ behaviour.

The disk amplitudes of one massless closed string mode and 
two open string modes were previously given in~\cite{kiem,garousi}. 
In particular in~\cite{garousi} (see also~\cite{garousi1})
the low energy effective action in 
the limit~\eqref{limit} was found and compared with the Born-Infeld action. 
In this paper, with the role of an open Wilson line in mind, we 
show that the effective action  can be written in a simpler and physically 
more transparent way than presented in~\cite{kiem,garousi}. 
The result also confirms the integration prescription~\eqref{giop}   
proposed in~\cite{liu,dastrivedi}. 

Physically, the appearance of the Wilson line and the integration
prescription can be understood from the ``stretched string'' effect
discussed in~\cite{lm}. Recall that when $B=0$, the low energy
external open string modes are point particles and the spacetime image 
of the worldsheet boundary is a single point, thus 
yielding local couplings between the closed string and gauge theory modes. 
With a nonzero $B$-field, the low energy open string behaves like an
electric dipole that is
in the presence of a strong background magnetic
field~\cite{bs,sj,yin}. As a result,
an open string mode (even on-shell) has a finite spacetime extension
$\Delta x^{\mu} = \theta^{\mu \nu} k_{\nu}$~\cite{bs,lm} proportional 
to its center of mass momentum. These open strings interact by
splitting and joining their ends.
When a closed string mode scatters off open string modes on a D-brane, 
all external open string modes join together to form a macroscopic 
open Wilson line, which then couples to the closed string mode. 
The extension of the Wilson line is given by
\begin{equation} 
\Delta x^{\mu} = \sum_a \Delta x_a^{\mu} = \theta^{\mu \nu} \sum_{a} k_{a \nu}
= - \theta^{\mu \nu} q_{\nu} \ ,
\end{equation}
where $k_a$ and $q$  are the momenta of open and closed string modes
respectively. From the worldsheet point of view, the Wilson line can 
then be viewed as the spacetime image of the worldsheet boundary,
and the integration procedure has its origin in the integrations of 
the vertex operator insertions on the boundary worldsheet. 
It is interesting that in the noncommutative theory,
the spacetime image of the worldsheet boundary, 
which is closed, is an open contour instead of a closed loop.

The way in which we write the string amplitudes previously given
in~\cite{kiem,garousi} makes it easy to extract the Wilson line from
the operators $\OO(k)$.
We will also see that the effective couplings illustrate to all orders the 
expectation that when multiple D-branes interact with background 
supergravity fields, the supergravity 
fields should be regarded not just as functions of the
$(p+1)$-dimensional spacetime but also as functionals of the nonabelian 
scalar fields (see e.g.~\cite{garousimyers,myers,kabat,taylor,garousi1}). 
For example, the coupling of a $Dp$-brane to the graviton can be written as
\begin{equation} \label{mye}
\int d^{p+1} x \;  {\rm STr} \left[h_{\mu \nu} (X(x), x) \, 
T^{\mu\nu} (x) \right]
\end{equation}
where $X$ are the nonabelian adjoint scalar fields of the D-branes and
$h_{\mu\nu}$ is the perturbation of the graviton.%
\footnote{We use the coupling~\eqref{mye}
to define the energy-momentum tensor, $T^{\mu\nu}$, of the
noncommutative theory; this
seems to differ from the Noether definition~\cite{noether}.}  
We also confirm to all orders the ``symmetrized trace'' prescription for the 
ordering in~\eqref{mye} up to terms quadratic in the field strength in 
$T^{\mu \nu}$.

This paper is organized as follows.  In section~\ref{sec:calc12} we
present the string amplitudes for the interaction of a closed string
with one or two open strings.  From this result, we extrapolate, in
section~\ref{sec:s} an action which involves the open Wilson line.
In section~\ref{sec:wl} we prove the proposal of section~\ref{sec:wl} 
by extracting  the contribution to the Wilson line
from the infinite series of disk diagrams with one closed string and
an arbitrary number of open strings.  This confirms the presence of
the open Wilson line, in the postulated form.  We conclude in 
section~\ref{sec:conc} with a discussion of the relation to Born-Infeld 
action. 

As this work was being completed, ref.~\cite{ooguri} appeared.  That paper
overlaps with this one. In particular, their energy momentum tensor 
agrees with the large $B$ limit of ours.

\section{Disk amplitudes with one closed string insertion} \label{sec:calc12}

\subsection{Preliminaries}
We will use $M,N = 0,1,\dots,9$ to denote the spacetime indices;
$\mu, \nu = 0,1,\dots,p$ to denote the worldvolume directions of 
a D-brane; and $i,j=p+1,\dots,9$ are the directions transverse to the brane.
The relations~\cite{sw} between
the closed (open) string metric $g$ ($G$) and coupling $g_s$ ($G_s$),
the $B$-field, and the noncommutativity parameter $\theta$ are
\begin{align} \label{ocmetric}
G^{-1} + \frac{\theta}{2 \pi \apr} & = \frac{1}{g + B}  \\
\intertext{and}
\label{occoup}
G_s & = g_s \left( \frac{\det G}{\det g} \right)^{\frac{1}{4}}.
\end{align}
We assume
that $B$ lies only in Neumann
directions 
(that is, along the $D$-brane)
and
that $g_{M N}$
vanishes for mixed Neumann/Dirichlet directions.
Therefore the above equations
apply to all spacetime directions; in particular, $G_{ij} = g_{ij}$ and 
$\theta^{ij} = 0$.

The NSNS vertex operators are given by
\begin{align}
\label{vclose}
V^{-1, -1}_{NSNS} & = \frac{2 g_c}{\apr} e^{- \phi - \tilde{\phi}}
e_{M N} \, \psi^{M} \tilde{\psi}^{N} e^{i q \cdot X}   \\
\label{vopen}
V^0_O  & =
a_{M} \, (i \dot{X}^{M}
+ 4 k \cdot \Psi \Psi^{M} ) e^{i k \cdot X} 
\end{align}
where the overdot in equation~\eqref{vopen} is understood to be
the tangential derivative along the worldsheet boundary 
for Neumann directions, and the normal derivative  for 
Dirichlet directions. 
The gauge bosons have momenta $k_a$ and polarizations $a_a$, with $a,b,\dots$
labeling the gauge boson%
\footnote{We have absorbed a factor of $g_{YM}$ into the
polarization.  Also, we will suppress Chan-Paton factors from the formulas
for simplicity, although we will comment on them when appropriate.}; the 
massless closed string mode has momentum $q$ and polarization $e_{M N}$.
We can decompose the closed string polarization into modes
corresponding to perturbations of the graviton, $h_{MN}$, the antisymmetric
$B$-field $b_{MN}$ and the dilaton $\varphi$ via
\begin{equation} \label{whatise}
e_{MN} = h_{MN} + b_{MN} + \half \varphi (g_{MN}-q_M \rho_N - q_N \rho_M).
\end{equation}
In addition, $h_{MN}$ is traceless---$h_{MN} g^{MN}=0$---and both
$h_{MN}$ and $b_{MN}$ are transverse: $h_{MN} g^{NP} q_P = 0 = b_{MN}
g^{NP} q_P$.
Note that we exclusively use the closed string metric when discussing
closed string fields.  The auxiliary vector $\rho_M$ obeys $q_M g^{MN}
\rho_N = 1$ and is introduced to
enforce transversality of $e_{MN}$,
\begin{equation} \label{qdote}
q_M g^{MN} e_{NP} = 0 = e_{PN} g^{NM} q_M,
\end{equation}
but it will not appear in any total amplitudes.  Also,
momentum conservation requires $q_\parallel = -k\equiv -\sum_a k_a$, where 
$q_{\parallel}$ and $q_{\perp}$  denote the components of $q$ 
parallel and perpendicular to the brane.
In addition to~\eqref{qdote}, the
on-shell conditions require that $q_{M} g^{M N} q_N = 0 = k_{a\mu}
G^{\mu\nu} k_{a\nu}$ and $k_{a \mu} G^{\mu \nu} a_{a \nu} = 0$.

We shall take the worldsheet to be the upper half plane 
and use the doubling trick to relate the the fields in 
the anti-holomorphic sector to those in the holomorphic sector
\begin{align} \label{mnd}
\tilde{X}^{\mu} (\bar{z}) & = \left(\frac{1}{g-B} (g+B) \right)^{\mu}{_\nu}
X^{\nu} (\bar{z}) \\
\tilde{X}^{i} (\bar{z}) & = - X^{i} (\bar{z})
\end{align}
and
\begin{align}
\tilde{\psi}^{\mu} (\bar{z})  &=  
\left(\frac{1}{g-B} (g+B) \right)^{\mu}{_\nu} \psi^{\nu} (\bar{z}), 
& \Psi^{\mu}  &= \half (\psi^{\mu} + \tilde{\psi}^{\nu}) =
\left(\frac{1}{g-B} g \right)^{\mu}{_\nu} \psi^{\nu},
 \\
\tilde{\psi}^{i}  &=  - \psi^{i}, 
&\Psi^{i}  &= \half (\psi^{i} - \tilde{\psi}^{i}) = \psi^{i}.
\end{align}
$\Psi^M$ is the open string supersymmetric partner of $X^M$ which
lives on the worldsheet boundary.
By introducing $D = \diag (1,1,\dots,-1,\dots,-1)$---an identity
matrix in the Neumann directions and minus the identity matrix
in the Dirichlet directions---the above equations can be written
collectively as
\begin{align}
\tilde{X}^{M} (\bar{z}) & = \left(\frac{1}{g-B} (g+B) D \right)^{M}{_N}
X^{N} (\bar{z}), \\
\tilde{\psi}^{M} (\bar{z}) & =  
\left(\frac{1}{g-B} (g+B) D \right)^{M}{_N} \psi^{N} (\bar{z}), \\
\Psi^{M}  & = \left(\frac{1}{g-B} g \right)^{M}{_N} \psi^{N} 
\end{align}

Note that in equations~\eqref{vclose} and~\eqref{vopen}
\begin{equation}
g_{YM}^2  = (2 \pi)^{(p-2)} G_s \apr^{\frac{p-3}{2}}, \qquad
g_c = \frac{\kappa_{10}}{2 \pi} = \sqrt{\pi} g_s^2 (2 \pi \apr)^2,
\end{equation}
where the open and closed string couplings $G_s$ and $g_s$ are related 
by~\eqref{occoup}. 
The overall normalization constant for the disk amplitudes is given 
by~\cite{jp}
\begin{equation}
C_{D_2} = \frac{1}{2 g_{YM}^2 \apr^2} \sqrt{\det G}
\end{equation}
and the brane tension is related to the Yang-Mills coupling constant $g_{YM}$ 
by
\begin{equation}
T_p = \frac{1}{(2 \pi)^{p} g_s \apr^{\frac{p+1}{2}}},\qquad
T_p \sqrt{\det (g + B)} = \frac{1}{g_{YM}^2 (2 \pi \apr)^2 }\sqrt{\det G}.
\end{equation}

The Green functions for the worldsheet fermions are%
\footnote{We are being a little sloppy in writing $\frac{g-B}{g+B}$
in~\eqref{ptp} (and subsequent equations)
and not specifying $\frac{1}{g}(g-B)\frac{1}{g+B}$ or
$\frac{1}{g+B}(g-B)\frac{1}{g}$ because, in fact, they are equal.}
\begin{align}
\vev{\psi^{M} (z) \psi^{N} (w)}  & = \frac{\apr}{2} \frac{g^{MN}}{z-w}
\\
\label{ptp}
\vev{\psi^{M}(z)  \tilde{\psi}^{N} (\bar{w})} & =
\frac{\apr}{2} \left(\frac{g-B}{g+B}D \right)^{MN}
\frac{1}{z-\bar{w}} \\
\vev{\psi^{M}(z) \Psi^{N} (w)} & =
\frac{\apr}{2} \left(\frac{1}{g+B}\right)^{MN}
\frac{1}{z-w} \\
\vev{\tilde{\psi}^{M} (\bar{z}) \Psi^{N} (w) } & =
\frac{\apr}{2} \left(\frac{1}{g-B} D \right)^{MN}
\frac{1}{\bar{z}-w} \\
\vev{\Psi^{M}(z) \Psi^{N} (w)} & =
\frac{\apr}{2} G^{MN}
\frac{1}{z-w}.
\end{align}

The Green functions for $X$ are given by
\begin{gather}
\begin{split}
\vev{ X^{\mu} (z, \bar{z})\; X^{\nu} (w, \bar{w})}
= - \apr &\left[ g^{\mu \nu} (\log\abs{z-w} - \log\abs{z- \bar{w}})
+ G^{\mu \nu} \log \abs{z - \bar{w}}^2 
\right. \\ & \left.
+ \frac{1}{2 \pi \apr} \theta^{\mu \nu} \log {\frac{z - \bar{w}}{\bar{z}-w}}
\right]
\end{split} \\
\vev{ X^{i} (z, \bar{z})\; X^{j} (w, \bar{w})}
= - \apr g^{i j} (\log\abs{z-w} - \log \abs{z- \bar{w}}).
\end{gather}
Using the above equations we can work out some basic formulas like 
\begin{equation} \label{corr1} \raisetag{3\baselineskip}
\begin{split}
A_n  & =  \vev{e^{i q \cdot X} (i) 
\prod_{a=1}^{n} e^{i k_a \cdot X} (y_a)} \\
& =  i \, 2^{\apr k^2} C_{D_2}^{X} (2 \pi)^d 
\delta(\sum_a k_a + q_{\parallel})
\prod_{a< b} 
\abs{\sin (\pi \tau_{ab})}^{2 \apr k_a \cdot k_b} \exp\left[\frac{i}{2} 
\bigl(k_a \times k_b\bigr) \bigl(2 \tau_{ab} -
\epsilon(\tau_{ab})\bigr)
\right] 
\end{split}
\end{equation}
where the $y_a$s are on the real axis and  $k = \sum_b k_b$.
For later convenience we have expressed
the above correlator in terms of $\tau_i$ which are defined by 
$y_a = - \cot (\pi \tau_a)$. Note that $0 \leq \tau_a \leq 1$ 
follows from $-\infty<y_a<\infty$.

We will also use 
\begin{equation} \label{corr2}
B_n  = a_M   \vev{e^{i q \cdot X} (i) \;
 i \dot{X}^M  (y_a) \prod_{b=1}^{n} e^{i k_b \cdot X} (y_b)} 
=  a_M V^M (y_a) A_n 
\end{equation}
with
\begin{multline} \label{bn}
a_M V^M (y_a) = 2 \apr \left[- (a \cdot k)\frac{y_a}{1+y_a^2}
+ \frac{i}{2 \pi \apr} (a \times k) \frac{1}{1+y_a^2}
+ i a_i g^{ij} q_{\perp j} \frac{1}{1+y_a^2} 
\right. \\ \left. + 
\sum_{b=1,b\neq a}^{n} (a \cdot k_b)\frac{1}{y_a-y_b} \right]
\end{multline}
where again $k = \sum_b k_b = - q_{\parallel}$. 
In this paper, when not specified otherwise, 
the dot product is with respect  to the open string metric and the 
cross product  denotes contraction using $\theta^{\mu\nu}$, i.e. 
$a \times k = a_{\mu} \theta^{\mu \nu} k_{\nu}$.

\subsection{One closed and one open}

The amplitude for the interaction of one massless closed string mode 
and one massless gauge field mode is given by 
\begin{equation} \label{amp1}
\begin{split}
{\mathcal A} & =  \vev{V^{-1,-1}_{NSNS} (i) \; V^0_O (0)} \\
& =  \frac{2 g_c}{\apr} e_{MN}\,  {\rm Tr} a_{P} \;
\vev{\psi^{M} \tilde{\psi}^{N} e^{i q \cdot X} (i)   
\, (i \dot{X}^{P} + 4 k \cdot \Psi \Psi^{P} ) e^{i k \cdot X} (0)}
\end{split}
\end{equation}
We have suppressed the ghosts, the relevant contribution from which
turns out to be unity.
In the following we shall
take the polarization of the closed 
string mode to be along the brane%
\footnote{An unfortunate by-product is that we will miss some couplings
of the dilaton to the transverse scalar fields and so we will 
not be able to see explicitly that $\rho_M$ of~\eqref{whatise} drops 
out in terms that involve scalar fields.}, i.e.\ $e_{MN} = e_{\mu \nu}$,
and we shall always omit the momentum conservation 
factor $(2 \pi)^{p+1} \delta (q_{\parallel} + \sum_a k_a)$. 
The amplitude~\eqref{amp1} can be evaluated using 
equations~\eqref{corr1} and~\eqref{corr2} to yield
\begin{equation} \label{oneone}
{\mathcal A} = \frac{\kappa_{10}}{g_{YM}^2 ( 2 \pi \apr)^2} 
{\rm Tr} \left[\frac{i}{2} \, E \, M - (2 \pi \apr) 
E^{\mu \nu} f_{\mu \nu} \right]
\skipthis{
{\mathcal A} = \frac{\kappa_{10}}{g_{YM}^2 ( 2 \pi \apr)^2} 
\left[\frac{i}{2} e_{M N} \left(\frac{g-B}{g+B} D \right)^{M N} \, M
+ (2 \pi \apr) T \right]
}
\end{equation}
where 
\begin{align} \label{defE}
E &=  e_{MN} \left(\frac{g-B}{g+B} D \right)^{M N}, &
E^{M N} &= e_{P  Q} \left(\frac{1}{g+B}\right)^{P M}
\left(\frac{1}{g- B } D \right)^{Q N}
\\
\intertext{and}
\label{defM}
M &= \theta^{\mu \nu} a_{\mu} k_{\nu} 
+  (2 \pi \apr) \phi_i g^{ij} q_{\perp j}, &
f_{\mu \nu} &=  i (k_{\mu} a_{\nu} -k_{\nu} a_{\mu}),
\end{align}
with $\phi_i \equiv a_i$.
\skipthis{
and 
\begin{equation}
\begin{split}
T & = i e_{M N} \, k_{\lambda} \xi^a_{P} T^a  \left[
\left(\frac{1}{g+B}\right)^{M P}
\left(\frac{1}{g-B} D \right)^{N \lambda} - 
\left(\frac{1}{g+B}\right)^{M \lambda}
\left(\frac{1}{g-B} D \right)^{N P} \right] \\
& = E^{\mu \nu} f_{\mu \nu} - E^{i \mu}
D_{\mu} \phi_i 
\end{split}
\end{equation}
with 
\begin{equation}
E^{\mu \nu} = e_{\lambda \rho} \left(\frac{1}{g+B}\right)^{\mu\lambda}
\left(\frac{1}{g- B}\right)^{\nu \rho},
\;\;\;\;\;\;
E^{i \mu} = g^{ij} \left[ e_{j \lam} (\frac{1}{g-B})^{\lambda \mu}
+ e_{\lambda j} (\frac{1}{g+B})^{\lambda \mu} \right]
\end{equation}
\begin{equation}
f_{\mu \nu} = - i (k_{\mu} A_{\nu} -k_{\nu} A_{\mu}),
 \,\,\;\;\;\;\;
D_{\mu} \phi_i  = -i k_{\mu} \phi_{i}
\end{equation}
}

\subsection{One closed and two open}

The amplitude with one massless closed string mode polarized along the 
brane direction and two noncommutative gauge modes is 
\begin{equation} \raisetag{\baselineskip}
\begin{split}
{\mathcal A} & =  \int_{- \infty}^{\infty} dy \, 
\vev{V^{-1,-1}_{NSNS} (i) \;\; V^0_O (0) \,\, V_O^0 (y)} \\
& =  \frac{2 g_c}{\apr} e_{M N} \, {\rm Tr} \, a_{1P} a_{2Q} 
\\ & \quad \times
\int \! dy 
\vev{\psi^{M} \tilde{\psi}^{N} e^{i q \cdot X} (i)   
\, (i \dot{X}^{P} + 4 k_1 \cdot \Psi \Psi^{P} ) e^{i k_1 \cdot X} (0)
\, (i \dot{X}^{Q} + 4 k_2 \cdot \Psi \Psi^{Q} ) e^{i k_2 \cdot X} (y)} \\
& =  \hat{{\cal A}} + \tilde{{\cal A}}_1 + \tilde{{\cal A}}_2  
\end{split}
\end{equation}
For convenience, we have 
split the amplitude into three parts:
$\hat{{\cal A}}$ contains the part involving the self-contraction of 
$\psi$ and $\tilde{\psi}$ in the closed string vertex operator;
$\tilde{{\cal A}}_1$ and $\tilde{{\cal A}}_2$ contain the remaining terms,
which respectively
involve four-fermion and six-fermion contractions.

Using~\eqref{corr1} and~\eqref{corr2} and integrating over $y$, the 
final result can be written as
\begin{align}
\label{ahat}
\hat{{\cal A}} &=  \frac{\kappa_{10}}{2 g_{YM}^2} \, E \, J \left[
(k_1 \cdot a_2)(k_2 \cdot a_1) - (a_1 \cdot a_2)(k_1 \cdot k_2)
- \frac{1}{(2 \pi \apr)^2} M_1 M_2 \right] \\*
& \qquad
 - \frac{\kappa_{10}}{8 \pi \apr g_{YM}^2 } E J \frac{2 \beta}{\apr t} 
\bigl[ (a_1 \cdot k_2) M_2 - (a_2 \cdot k_1) M_1 + (k_1 \times k_2)
(a_1 \cdot a_2) \bigr]
\nonumber \\
\label{atilde1}
\tilde{{\cal A}}_1 &= - i \frac{\kappa_{10}}{g_{YM}^2}  J  E^{\mu \nu}
\left[ \frac{1}{2 \pi \apr} (f_{1\mu \nu}  M_2 + f_{2 \mu \nu} M_1) 
- \frac{\beta}{\apr t} 
\left(f_{1\mu \nu}  (a_2 \cdot k_1) - f_{2 \mu \nu}
(a_1 \cdot k_2) \right) \right]
\\
\label{atilde2}
\tilde{{\cal A}}_2 &= - \frac{\kappa_{10}}{g_{YM}^2}  J
E^{\mu \nu} \left[- f_{1 \mu \rho} f^{\rho}_{2 \nu} + 
\p_{\mu} \phi^i_1 \p_{\nu} \phi_{2 i} + (1 \rightarrow 2) 
\right. \nonumber \\* & \qquad \left. -
\frac{\beta}{\apr t} \left( - f_{1 \mu \rho} f^{\rho}_{2 \nu} + 
\p_{\mu} \phi^i_1 \p_{\nu} \phi_{2 i} - (1 \rightarrow 2) \right) \right]
\end{align}
where $E$ and $E^{\mu \nu}$ were  defined in 
equations~\eqref{defE} and~\eqref{defM}. $M_{1}$ and  $f_{1\mu
\nu}$  generalize~\eqref{defM} to
\begin{align} \label{defMi}
M_{1} &= \theta^{\mu \nu} a_{1\mu} k_{\nu} 
+  (2 \pi \apr) \phi_{1i} g^{ij} q_{\perp j}, &
f_{1\mu \nu} &=  i (k_{1\mu} a_{1\nu} -k_{1\nu}
a_{1\mu}),
\end{align}
and similarly for $M_2, f_{2\mu\nu}$.
Note that the {\em total} (open string) momentum $k$ appears in the
definition of $M_{1,2}$; this will be important when we relate terms
involving $M$ to the Wilson line.
We have also introduced
\begin{gather}
\begin{align}
\beta &= \frac{k_1 \times k_2}{\pi}, &
t &= - k^2  = - 2 k_1 \cdot k_2, &
\p_{\mu} \phi_{1i} &= i k_{1 \mu} \phi_{1i},  
\end{align} \\
\label{expJ}
J =  \frac{\Gamma(1 - \apr t)}{\Gamma\bigl(1- \frac{\apr t}{2} + \beta \bigr)
\Gamma\bigl(1-\frac{\apr t}{2} - \beta \bigr)} = 
\frac{\sin \frac{k_1 \times k_2}{2}}{\frac{k_1 \times k_2}{2}}
+ \order{\apr t} + \cdots ,
\end{gather}
In~\eqref{ahat}--\eqref{atilde2} the dot product is with respect to the 
open string metric%
\footnote{In particular $a_1 \cdot a_2 = a_{1\mu} G^{\mu \nu} a_{2 \nu}
+ a_{1i} G^{ij} a_{2j} = a_{1\mu} G^{\mu \nu} a_{2 \nu} +
\phi_{1i} g^{ij} \phi_{2j}$.}
and the indices are raised and lowered by the 
open string metric.

We shall be interested in the leading terms of the amplitude 
in an $\apr t$ expansion i.e.\ we consider the limit 
$\apr k_{\mu} G^{\mu \nu} k_{\nu} \ll 1$, while keeping 
$k_{1\mu} \theta^{\mu \nu} k_{2 \nu}$ finite.  Substituting the
leading term in the expansion~\eqref{expJ}
into~\eqref{ahat}--\eqref{atilde2}, the amplitude
separates
into finite terms and terms containing poles in  $t$.
The finite terms are
\begin{align} \label{contac1}
\hat{{\cal A}}_f &= \frac{\kappa_{10}}{2 g_{YM}^2} E 
\frac{\sin \frac{k_1 \times k_2}{2}}{\frac{k_1 \times k_2}{2}}\left[
- \frac{1}{(2 \pi \apr)^2} M_1 M_2 + \half f_{1 \mu \nu} f_2^{\mu \nu}
+ G^{\mu \nu} g^{ij} \p_{\mu} \phi_{1i} \p_{\nu} \phi_{2j} \right]
\\ \label{contac2}
\tilde{{\cal A}}_{1f} &=  i \frac{\kappa_{10}}{g_{YM}^2}  
\frac{\sin \frac{k_1 \times k_2}{2}}{\frac{k_1 \times k_2}{2}}
\frac{1}{2 \pi \apr} E^{\mu \nu} (f_{1\mu \nu}  M_2 + f_{2 \mu \nu} M_1)
\\ \label{contac3}
\tilde{{\cal A}}_{2f} &= - \frac{\kappa_{10}}{g_{YM}^2}  
\frac{ \sin \frac{k_1 \times k_2}{2}}{\frac{k_1 \times k_2}{2}}
E^{\mu \nu} \left[- f_{1 \mu \rho} f^{\rho}_{2 \nu} + 
\p_{\mu} \phi^i_1 \p_{\nu} \phi_{2 i} + (1 \rightarrow 2) \right]
\nonumber \\* & \qquad
+  \frac{\kappa_{10}}{\pi \apr g_{YM}^2}  \sin \frac{k_1 \times k_2}{2} 
E^{\mu \nu} (a_{1\mu} a_{2 \nu} - a_{2\mu} a_{1 \nu}),
\end{align}
and the terms with a pole are
\begin{align}
\hat{{\cal A}}_p &= - \frac{\kappa_{10}}{2 \pi^2 \apr  g_{YM}^2 } E
\frac{\sin \frac{k_1 \times k_2}{2}}{\apr  t}
\bigl[ (a_1 \cdot k_2) M_2 - (a_2 \cdot k_1) M_1 + (k_1 \times k_2)
(a_1 \cdot a_2) \bigr] \\
\tilde{{\cal A}}_{1p} &=  \frac{2i}{\pi} 
\frac{\kappa_{10}}{g_{YM}^2 }
\frac{\sin \frac{k_1 \times k_2}{2}}{\apr  t} E^{\mu \nu}
\left[f_{1 \mu \nu} (a_2 \cdot k_1 ) - f_{2 \mu \nu} (a_1 \cdot k_2)
\right] \\
\tilde{{\cal A}}_{2p} &= \frac{2}{\pi} 
\frac{\kappa_{10}}{g_{YM}^2 }
\frac{\sin \frac{k_1 \times k_2}{2}}{\apr  t} E^{\mu \nu}
\left[ - (f_{1 \mu \rho} f_2^{\rho}{_\nu})_p + 
\p_{\mu} \phi^i_1 \p_{\nu} \phi_{2 i} - (1 \rightarrow 2) \right]
\end{align}
where 
\begin{equation}
\left[f_{1 \mu \lam} f_2^{\lam}{_\nu}\right]_p
= -  \left[ (a_2 \cdot k_1) a_{1 \mu} k_{2 \nu}
+  (a_1 \cdot k_2) a_{2 \nu} k_{1 \mu} -
(a_1 \cdot a_2) k_{1 \mu} k_{2 \nu} \right]
\end{equation}
Note that the fourth term in $f_{1 \mu \lam} f_2^{\lam}{_\nu}$
contributes to the finite term.
The pole terms are all proportional to $\sin \frac{k_1 \times k_2}{2}$
and can be reproduced from exchanging massless gauge bosons (or 
scalar fields) using the vertex~\eqref{oneone}  and cubic 
vertices of gauge bosons, as was pointed out in~\cite{garousi}.

The finite terms in~\eqref{contac1}--\eqref{contac3} are all proportional 
to $\frac{\sin \frac{k_1 \times k_2}{2}}{\frac{k_1 \times k_2}{2}}$,
except for
the last term in~\eqref{contac3} which is proportional to 
$\sin \frac{k_1 \times k_2}{2}$. The factor 
$\frac{\sin \frac{k_1 \times k_2}{2}}{\frac{k_1 \times k_2}{2}}$ gives
rise to the generalized star-product
$\ast_2$ between two Yang-Mills fields~\cite{garousi}, which in momentum
space is
\begin{equation} \label{ast2}
f(k_1) \ast_2 g(k_2) = 
f(k_1) \frac{\sin \frac{k_1 \times k_2}{2}}{\frac{k_1 \times k_2}{2}} g(k_2) 
\end{equation}
The last term of~\eqref{contac3} combines with the second 
term in~\eqref{oneone} to give the noncommutative field strength 
\begin{equation} \label{ncf}
F_{\mu \nu} = \del_{\mu} A_{\nu} -  \del_{\nu} A_{\mu}
 - i A_{\mu} \ast A_{\nu} + i A_{\nu} \ast A_{\mu}
\end{equation}
which in momentum space becomes 
\begin{equation} \label{fncf}
F_{\mu \nu} = i (k_{\mu} A_{\nu} - k_{\nu} A_{\mu}) - 
2 \sin \frac{k_1 \times k_2}{2} A_{\mu} (k_1) A_{\nu} (k_2)
\end{equation}

\section{Supergravity Couplings of Noncommutative D-branes} \label{sec:s}

From equations~\eqref{oneone} and~\eqref{contac1}--\eqref{contac3}
we are ready to write down the effective couplings of the massless NSNS 
closed string modes to the noncommutative D-brane up to terms 
quadratic in Yang-Mills fields. Since $M$ defined in~\eqref{defM} 
and~\eqref{defMi} involve the total momentum of the open string modes it is 
more convenient to write down the couplings in momentum space%
\footnote{We have included a tadpole term which can be easily
obtained, for example, by cutting open the annulus diagram.} 
\begin{smaleq}
\begin{multline} \label{preeff}
S = \frac{\kappa_{10}}{g_{YM}^2 (2 \pi \apr)^2} 
\int \frac{d^{10} q}{(2 \pi)^{10} } \frac{d^{p+1}k_1}{(2 \pi)^{p+1} }
 \frac{d^{p+1}k_2}{(2 \pi)^{p+1} } \, 
\sqrt{\det G} \, (2 \pi)^{p+1} \delta^{(p+1)} (q_{\parallel}+k_1 + k_2) \\
\times \biggl\{ \half E (q)  
\left[1 + iM (k_1) - \half M (k_1) {\ast_2} M (k_2) 
+ \frac{1}{4} 
\sF_{\mu \nu} (k_1){\ast_2} \sF^{\mu \nu} (k_2)
  + \frac{1}{2} 
D_{\mu} X_i (k_1) {\ast_2} D^{\mu} X^i (k_2)\right] \\
+ \ha E^{\mu \nu} (q) \left[ - \sF_{\mu \nu} (k_1) 
- i  \sF_{\mu \nu}(k_1) {\ast_2} M (k_2)  + 
\sF_{\mu\rho} (k_1) {\ast_2} \sF^{\rho}{_\nu} (k_2) 
- D_{\mu} X^i (k_1) {\ast_2}  D_{\nu} X_i (k_2) \right] \biggr\}
\end{multline}
\end{smaleq}
where $\sF = \papr F$ is the noncommutative field strength~\eqref{ncf}, 
$X_i = 2 \pi \apr \phi_i$ and $D_{\mu} X_i  = 
\del_{\mu} X_i  - i [A_{\mu}, X_i ]_{\ast}$.   Note that we have
promoted the perturbation $a_M$ to the full gauge field $A_M$.  $\sF$ and 
$D X$ should be understood as their Fourier transforms, as
in~\eqref{fncf}.  For terms linear in Yang-Mills fields it is
understood that $k_2$ should be set to zero and not integrated, and
for the tadpole term,  it is understood that $k_1=k_2=0$.
Except for those in $E$ and $E^{\mu \nu}$ 
the indices are always raised and lowered using 
the open string metric (recall $G_{ij} = g_{ij}$). The
$\ast_2$-product was defined 
in~\eqref{ast2} and 
\begin{equation} \label{feffm}
M(k_i) = \theta^{\mu \nu} q_{\mu} A_{\nu}(k_i) + q_{\perp} \cdot X (k_i).
\end{equation}
For the graviton  polarized along the brane
\begin{equation} \label{e1}
E =  h_{\mu \nu} \left(\frac{g-B}{g+B}  \right)^{\mu \nu}, \qquad
E^{\mu \nu} = h_{\lambda \rho} \left(\frac{1}{g+B}\right)^{\lambda \mu}
\left(\frac{1}{g- B}\right)^{\rho \nu} \ .
\end{equation}
with $h_{\mu \nu}$  symmetric, 
traceless in terms of closed string metric, i.e. $g^{\mu \nu} h_{\mu \nu}
=0$. A similar expression applies to  the NSNS anti-symmetric tensor 
$b_{\mu \nu}$. Note $E$ and $E^{\mu \nu}$ in~\eqref{e1} satisfy the relation
\begin{equation} \label{interE}
E = G_{\mu \nu} E^{\nu \mu}
\end{equation}
For the dilaton 
\begin{equation} \label{e3}
E = \varphi  ( g_{\mu \nu} G^{\mu \nu} - 4), \qquad
E^{\mu \nu} = \ha \varphi \, g_{\lambda \rho} \left(\frac{1}{g+B}\right)^{\lambda \mu}
\left(\frac{1}{g- B}\right)^{\rho \nu} \ .
\end{equation}

It can be checked that the effective action~\eqref{preeff} is not gauge 
invariant  under the noncommutative gauge transformations
\begin{equation}\label{tran}
\begin{split}
\delta_{\lambda} A_\mu & = \partial_\mu \lambda + 
i \lambda \ast A_\mu - i A_\mu \ast \lambda \, , \\
\delta_{\lambda} X_i  & = 
i \lambda \ast X_i - i X_i \ast \lam
.
\end{split}
\end{equation}
This is almost the same situation encountered in the one-loop 
effective action of noncommutative super-Yang-Mills theory~\cite{lm2,zanon},
which is related to the disk amplitudes considered here by factorization.
The reason is by now well understood; the non-gauge-invariance 
of higher order terms in the action will conspire to cancel those 
from~\eqref{preeff} to make the whole effective action gauge 
invariant~\cite{mehen,liu,zanon1}.
Using the result  of~\cite{liu} regarding  the relation between 
the $\ast_n$ products and an open Wilson line, the resolution 
is to attach each $\sF$ and $X$ in~\eqref{preeff} to a straight Wilson line
and integrate over the insertion positions with respect to the 
path ordering.  The presence of the Wilson line can already been 
seen from the nature of the $M$-dependence in~\eqref{preeff}. 
Thus we obtain the following gauge 
invariant completion of the above action~\eqref{preeff}
\begin{subequations} \label{effa}
\begin{equation} \label{effe} 
S  = \frac{\kappa_{10}}{g_{YM}^2 (2 \pi \apr)^2} 
\int \frac{d^{10} q}{(2 \pi)^{10}} \, 
\sqrt{\det G} \,\; \half \; E^{\mu \nu} (q) \, T_{\mu \nu} (- q)
\end{equation} 
where 
\begin{equation} \label{efft}
\begin{split} 
T_{\mu \nu}( - q)  & =  {\rm Tr} \int d^{p+1} x  \,
L_{\ast} 
\left( W(x,C) \; T_{\mu \nu} (x) \right) \ast e^{i q_{\rho} x^{\rho}} \\
T_{\mu \nu} (x) & = G_{\mu \nu} 
\left(  1 + \frac{1}{4} \sF_{\lam \rho} \sF^{\lam \rho} + 
\frac{1}{2} 
D_{\lam} X_i D^{\lam} X^i \right) 
 - \sF_{\mu \nu}
+ \sF_{\mu\rho} \sF^{\rho}{_\nu} - D_{\mu} X^i  D_{\nu} 
X_i 
\end{split}
\end{equation}
\skipthis{
& \biggl[G_{\nu \mu} 
\left(  1 + \frac{1}{4} \sF_{\lam \rho} \sF^{\lam \rho} + 
\frac{1}{2} 
D_{\lam} X_i D^{\lam} X^i \right) \\
& + \sF_{\mu \nu}
+ \sF_{\mu\rho} \sF^{\rho}_{\; \nu} - D_{\mu} X^i  D_{\nu} 
X_i  \biggr] \biggr\} \ast e^{i q_{\mu} x^{\mu}} 
\end{split}
\end{equation}
}%
\skipthis{
\, \biggl\{ \half E \; {\rm STr} \left[e^{i M} \left(
1 + \frac{1}{4} 
F_{\mu \nu} F^{\mu \nu} + 
\frac{1}{2} 
D_{\mu} \phi_i D^{\mu} \phi^i \right) \right] \\
& + \ha E^{\mu \nu} \; {\rm STr} \left[e^{i M} \left( F_{\mu \nu}
+ F_{\mu\rho} F^{\rho}_{\, \nu} - D_{\mu} \phi^i  D_{\nu} 
\phi_i \right) \right]\biggr\}
\end{split}
\end{equation}
}%
and ($\xi^{\nu} (\tau) = \theta^{\mu \nu} q_{\mu} \tau$)
\begin{equation} \label{cwilson}
W(x,C)  =  P_{\ast} \exp \left[ i 
\int_0^1 d \tau  \,
\left(q_{\mu} \theta^{\mu \nu} A_{\nu} (x + \xi (\tau))
+ q_{\perp i}  X^i ( x + \xi (\tau)) \right) \right].
\end{equation}
\end{subequations}
Clearly,~\eqref{cwilson} leads to the expansion in $M$ seen in~\eqref{preeff}.
In~\eqref{efft} we have used a short hand notation $L_{\ast}$ 
to denote the procedure of integrating each $\sF$ or $X$ in $T_{\mu \nu}$ 
along the Wilson line via path ordering.%
\footnote{For example,
\begin{multline*} \label{giope}
\int d^{p+1} x  \, L_{\ast}  
\left[ W(x,C) \;  F_{\mu \rho} (x) F^{\rho}_{\;\,\nu} (x) \right] 
\ast e^{i q_{\mu} x^{\mu}} \\
=  \int d^{p+1} x  \!  \int^{1}_{0}\! d \tau_1 \int_{0}^1 \! d \tau_2 \; 
P_{\ast} \left[ W(x,C) 
F_{\mu \rho} (x + \xi(\tau_1)) F^{\rho}{_\nu} (x + \xi(\tau_2))
 \right] \ast e^{i q \cdot x} .
\end{multline*}} 

In general, 
\begin{equation} \label{wils}
\begin{split}
&    \int d^{p+1} x  \,
 L_{\ast} \left[ W(x,C) \prod_{a=1}^n \OO_a (x)
 \right] \ast e^{i q \cdot x}  \\
& \equiv \int d^{p+1} x  \! \left( \prod_{a=1}^{n}
 \int^{1}_{0}\! d \tau_a  \right) \; 
  P_{\ast} \left[ W(x,C) \prod_{a=1}^n \OO_a (x + \xi(\tau_a))
 \right] \ast e^{i q \cdot x} \\
& = \sum_{m=0}^{\infty} 
\frac{i^m}{m!}  \left( \prod_{i=a}^{n+m} 
\int \! \frac{d^{p+1} k_a}{(2 \pi)^{p+1}} \right) \,
Q_m (k_1, \cdots, k_{n+m})
\end{split}
\end{equation}
where in the third line we have expanded the Wilson line in terms of
the power
series of the gauge fields
with ($\OO_i(k)$ are the Fourier transforms of $\OO_i(x)$)
\begin{smaleq}
\begin{equation} \label{qmm}
Q_m = (2 \pi)^{p+1} \delta^{(p+1)}(q + \sum_{i=1}^{m+n} k_i) \,  
\OO_1 (k_1) \cdots \OO_n (k_n) M(k_{n+1}) \cdots 
M(k_{n+m}) \,  J_{n+m} (k_1, \cdots, k_{n+m})
\end{equation}
\end{smaleq}
and 
\begin{equation} \label{defjn}
J_n (k_1, \cdots, k_n) = \int_{0}^{1} d \tau_1 \cdots 
\int_{0}^{1} d \tau_n \; \exp \left[  \frac{i}{2} \sum_{a<b}^n
(k_a \times k_b) (2 \tau_{ab} - \epsilon (\tau_{ab})) \right].
\end{equation}
It is important to note that all the entries in~\eqref{qmm} are completely 
symmetric under the ordering due to symmetric properties of the $J_{m+n}$.
For simplicity, we have written equations (3.10)--(3.12) as if the
gauge group were abelian. With Chan-Paton factors, we should include a
path ordering with respect to the $\tau_i$s when multiplying the
adjoint-valued objects; this orders the operators according to their
location on the Wilson line. This prescription clearly maintains the
full symmetry of $J_{m+n}$.

In the next section we shall show that higher order terms in~\eqref{effa} 
can  be systematically extracted from the amplitudes of one 
closed string with arbitrary number of open string modes. 
In particular, the structure of the amplitude is such that one obtains
a fully symmetrized action, though only in the low
energy limit.

The effective action~\eqref{effa} indicates that when we turn on a 
constant $B$-field, the D-brane becomes a source for the dilaton (even for
a $D3$-brane) and the NSNS $B$-field, in addition to the graviton. 
This is expected as we 
know that the supergravity 
solution for D3-branes with a $B$-field involves nontrivial dilaton and 
$B$-field background~\cite{hi,mr}. Since a constant $B$-field does not 
break supersymmetry, there will still be no net force between parallel
D-branes. Thus the 
exchange amplitude due to these additional couplings have to cancel 
with the new couplings in the R-R sector or among themselves. 
Similarly there are new couplings at the linear and quadratic level, and 
their contributions to $F^2$ and $F^3$ interactions between the branes 
should  cancel one another or with those in RR-sector since, 
as shown in~\cite{lm2}, there is no one-loop correction to 
$F^2$ and $F^3$ terms in ${\cal N} = 4$ noncommutative super-Yang-Mills 
theory. In~\cite{liu}, minimal couplings which reproduce
the one-loop $F^4$ terms were worked out and it was found for example that 
the graviton couples to noncommutative gauge field modes in a form   
\begin{equation} \label{coup}
h_{\lam \rho} (q) g^{\lam \mu} G^{\rho \nu} \,
\int d^{p+1} x  \, L_{\ast}  
\left[ W(x,C) \; \left(F_{\mu \sigma} F_{\nu}{^\sigma} - \frac{1}{4} 
 G_{\mu \nu} F_{\sigma \alpha} F^{\sigma \alpha} \right) \right] \ast
 e^{i q_{\rho} x^{\rho}} \ .
\end{equation}
The quadratic in field strength coupling to the graviton that we found
above is
\begin{smaleq}
\begin{equation} \label{coupe}
h_{\lam \rho} (q)   \left(\frac{1}{g+B}\right)^{\lambda \mu}
\left(\frac{1}{g- B}\right)^{\rho \nu}
\int d^{p+1} x  \, L_{\ast}  
\left[ W(x,C) \; \left(F_{\mu \sigma} F_{\nu}{^\sigma} - \frac{1}{4} 
 G_{\mu \nu} F_{\sigma \alpha} F^{\sigma \alpha} \right) \right] \ast
 e^{i q_{\rho} x^{\rho}}
\end{equation}
\end{smaleq}
The general structures of~\eqref{coup} and~\eqref{coupe} 
agree very well except for the index structure of the graviton. 
This is to be expected since~\eqref{coup} was inferred indirectly 
to reproduce the one-loop $F^4$ terms and this procedure 
does not determine the Lorentz structure uniquely. 

To gain some intuition about the physical meaning of~\eqref{effa} 
let us look at some special cases.

\subsection{$B=0$}

In this case $E^{\mu \nu}$ is the standard supergravity 
perturbation and equation~\eqref{efft} becomes
\begin{equation}
T_{\mu \nu} (-q) = \int d^{p+1} x \; e^{iq_{\perp i} X^i} \,  
e^{i q_{\parallel \rho} x^{\rho}} \, T_{\mu \nu} (x)
\end{equation} 
and thus equation~\eqref{effe} becomes
\begin{equation} 
\begin{split}
S & =  \frac{\kappa_{10}}{g_{YM}^2 (2 \pi \apr)^2} \int \frac{d^{10}
q}{(2 \pi)^{10}} \,
\sqrt{\det G} \,\; \half E^{\mu \nu} (q) \, \left[\int d^{p+1} x \, 
e^{iq_{\perp i} X^i} \,  e^{i q_{\parallel \rho} x^{\rho}}\,
 T_{\mu \nu} (x) \right] \\
& =  \frac{\kappa_{10}}{g_{YM}^2 (2 \pi \apr)^2}\int d^{p+1} x \,
\sqrt{\det G} \,\; \half \; E^{\mu \nu} (X(x),x) \; T_{\mu \nu} (x)
\end{split}
\end{equation}
where in the above we have used 
\begin{equation} \label{frou}
 E^{\mu \nu} (X(x),x) = \int \frac{d^{10} q}{(2 \pi)^{10}} \,
 E^{\mu \nu} (q) \,e^{iq_{\perp i} X^i} \,  e^{i q_{\parallel \rho} x^{\rho}}
\end{equation}
The above result indicated that the external supergravity fields
should be considered as a functional of the transverse scalar 
fields. 

In the nonabelian case, the $L_{\ast}$ in~\eqref{efft} reduces to a 
symmetrized trace and the action becomes:
\begin{equation} 
S =  \frac{\kappa_{10}}{g_{YM}^2 (2 \pi \apr)^2}\int d^{p+1} x \,
\sqrt{\det G} \,\; \half \;{\rm STr} \left[E^{\mu \nu} (X_i(x),x) \; 
T_{\mu \nu} (x) \right]
\end{equation}
Note that the  $X_i (x)$ are now nonabelian matrices and the symmetrized
trace  is to be understood as first doing the Fourier transform  
as in~\eqref{frou} and then treat the exponential as a power series.

\subsection{Relation with  Matrix Theory} \label{sec:mt}

In this section we shall consider the Euclidean version~\cite{ikkt} 
of Matrix Theory~\cite{bfss}.
Noncommutative Yang-Mills theory is obtained naturally from the 
$U(\infty)$ limit of  Matrix Theory~\cite{aoki,ishibashi,iikk,seiberg}. 
This is closely related to the construction of D$p$-Branes in
Matrix Theory (see e.g.~\cite{bss,miao,ishibashi}).
In terms of Matrix Theory variables $X^{M}$, D$p$-branes
can be  obtained as a classical solution
\begin{equation} \label{commu}
[x^{\mu}, x^{\nu}] = i \theta^{\mu \nu}, \qquad
\mu, \nu = 0,\cdots, p
\end{equation}
and the noncommutative Yang-Mills theory arises by considering the 
dynamics of the fluctuations around the classical solution
\begin{equation} \label{matrix}
X^{\mu} = x^{\mu} + \theta^{\mu \nu} A_{\nu} 
\end{equation} 
and 
\begin{equation} \label{matrf}
[X^{\mu}, X^{\nu}] = -i 
\left[\theta (F - \theta^{-1}) \theta \right]^{\mu \nu}
\end{equation}

As pointed out in~\cite{seiberg}, the noncommutative Yang-Mills theory 
obtained this way corresponds to the choice of $\Phi = -B$ description of 
the open string dynamics~\cite{sw}. In this description, 
the open string parameters given by
\begin{eqnarray}
\label{t1}
\theta & = & \frac{1}{B} \\
G^{-1} & = & - \frac{1}{B} g \frac{1}{B} \\
\label{t3}
G_s & = & g_s \sqrt{\det B g^{-1}} 
\end{eqnarray}
On the other hand the off-shell 
action~\eqref{effe}--\eqref{cwilson}, which was extrapolated 
from the on-shell string amplitudes, corresponds to the choice 
$\Phi = 0$. The two descriptions are in general 
related by  field redefinitions~\cite{sw}. 

In the large $B$-field limit, the mixed Neumann and Dirichlet 
boundary conditions~\eqref{mnd}
become effectively Dirichlet conditions and thus we expect that our 
results~\eqref{effe}--\eqref{cwilson} can be expressed naturally 
in terms of  Matrix Theory  variables $X$. Indeed it can be checked
that equations~\eqref{ocmetric} and~\eqref{occoup} reduce 
to~\eqref{t1}--\eqref{t3} in this limit.  
We note that the open Wilson line~\eqref{cwilson} can be written in
Matrix Theory variables~\eqref{matrix} as~\cite{wadia}
\begin{equation} 
W(x, C) = \exp \left(i q \cdot X \right) = \exp 
\left(i q_{\mu}  X^{\mu} + i q_i X^i \right)
\end{equation}
After some algebraic manipulations it can be shown that in the large 
$B$-field limit, the action~\eqref{effe}--\eqref{cwilson} can be 
rewritten  (e.g. for the graviton) as
\begin{equation}
S = \frac{\kappa_{10}}{(2 \pi)^{\frac{p-1}{2}} 
g_s \apr^{(\frac{p+1}{2})}} \; 
\int  \frac{d^{10} q}{(2 \pi)^{10}} \, 
\sqrt{\det \theta} \; h_{\mu \nu} (q) \; {\rm STr} \biggl[ e^{i q \cdot X} 
[ X^{\mu}, X^{M}] \,g_{MN} \, [X^{N}, X^{\nu}] \biggr]
\end{equation} 
where we have used~\eqref{matrf}. In the above ``Tr'' should be 
understood to be defined in the Hilbert space given by~\eqref{commu}, i.e.
\begin{equation}
\sqrt{\det \theta} \; {\rm Tr } = 
\int \frac{d^{p+1} x}{(2 \pi)^{\frac{p+1}{2}}}  {\rm tr}
\end{equation}
where here ``tr'' denotes the group trace,  
and the symmetrized trace ``STr'' should be understood as 
a formal way of writing the $L_{\ast}$ prescription defined 
below~\eqref{cwilson}.

\section{The structure of higher-point amplitudes 
and the open Wilson line} \label{sec:wl}

In this section we will show how, by considering disk amplitudes with
one massless closed string and an arbitrary number of massless open
string modes, one finds the full open Wilson line in the effective
action~\eqref{effa}. We will not attempt to evaluate the general 
amplitudes and shall be only interested in the parts of the 
amplitudes relevant for the higher order terms in~\eqref{effa}. 
As we have seen the last section, they should arise as {\it finite}
terms of the amplitudes in the limit $\apr k_a \cdot k_b \ra 0$.%
\footnote{We will ignore terms with a pole in $k_a \cdot k_b$ as they 
arise from exchanging massless gauge bosons (or scalars) using 
lower order vertices.} 
For the convenience of the discussion 
we shall not write explicitly the numerical prefactors of the 
amplitudes. 

The general disk amplitude is
\begin{equation} \label{namp}
\begin{split}
{\mathcal A}_m & =  
\left[\prod_{a=2}^{m} \int_{-\infty}^{\infty} dy_a \right] 
\vev{V^{-1,-1}_{NSNS} (i) \; 
\prod_{a=1}^{m} V_O^0 (\xi_a, k_a; y_a)} \\
& =   e_{M N}  \int \! dy 
\vev{\psi^{M} \tilde{\psi}^{N} e^{i q \cdot X} (i)   
\, \prod_{a=1}^m a_{a M_a} 
(i \dot{X}^{M_a} + 4 k_a \cdot \Psi \Psi^{M_a} ) e^{i k_1 \cdot X} (y_a)}.
\end{split}
\end{equation}
We may take $y_1=0$ using the $SL(2,{\mathbb{R}})$
invariance of the disk, and in the second line we have used a
shorthand for all the $y_a$-integrations.

There are two important features 
of the correlators inside the integrals which will be relevant for us.
There is a nontrivial phase factor
\begin{equation} \label{get*n}
\prod_{a< b} \exp \left[\frac{i}{2} 
(k_a \times k_b) (2 \tau_{ab} - \epsilon (\tau_{ab})) \right]
\end{equation} 
as was given in~\eqref{corr1}.  (Note that this just gives unity when
$B=0$.)  The $\epsilon(\tau_{ab})$ term is probably quite familiar by
now, from~\cite{sw} and much subsequent work.  It arises from bosonic
contractions between open string vertex operators.  Were it not for
the linear term, which arises from contractions between the open
strings and the closed string, the phase factor would give rise to
the $\ast$-product.%
\footnote{The linear term vanishes when there is no momentum
flow between the closed and open strings. The phase factor also appeared
in one-loop diagrams~\cite{lm2}; this is more than just a coincidence 
as the one-loop diagrams must factorize into tree diagrams.}
Equation~\eqref{get*n} is precisely 
the same phase factor that appears in the expansion of the open 
Wilson lines~\eqref{defjn}. 

Another general feature is that for
every $i\Dot{X}^M(y_a)$ insertion from an open string vertex operator,
there is a term of the form
\begin{equation} \label{getM}
i \left[(q \times a_a) + (2 \pi \apr) \phi_a \cdot q_{\perp} \right]
\frac{1}{\pi (1 + y_a^2)},
\end{equation}
coming from the contraction with $e^{i q \cdot X}$ in the closed string 
vertex operator, as was already given in~\eqref{bn}.
This, of course, is precisely the $iM_a$ of
~\eqref{feffm}.

Na\"{\i}vely,
the Wilson lines now can be seen to arise  as follows.
Suppose we are considering what is  an interaction between
$n$ open strings and a closed string.  For each additional open string
that we add, one of the contributions is an additional factor of
$iM_a$.  Because this comes from the
$i\Dot{X}^M(y_a)$ in the vertex operator, the rest of the amplitude
is almost unchanged from that without the additional open string; the
only other change is in the phase factor~\eqref{get*n} from the
additional $e^{ikX}$.  But
this is precisely the open
Wilson line expansion~\eqref{wils}--\eqref{defjn}.
In practice the story is more complicated as there are additional 
functions and singularities of of $y_a$ in the integrand.

As the simplest example let us look at the (tadpole) term 
\begin{equation}
E (q) \; \int d^{p+1} x \; W(x,C) \ast e^{i q_{\mu} x^{\mu}} 
\end{equation}
in~\eqref{effa}.
From what was outlined just after equation~\eqref{getM}, the relevant
terms are the
products of $iM$
coming from the contraction of $i \dot{X}$  with the $e^{i q \cdot X}$
part of the closed string vertex operator. Thus the relevant terms are  
\skipthis{
we look at the subset of terms in 
~\eqref{namp} which are given by:
\begin{equation} \label{cn}
\begin{split}
& \int_{- \infty}^{\infty} \prod_{i=2}^{m} dy_i \, 
\vev{\psi^{M}(i) \tilde{\psi}^{N} (-i)} \vev{ e^{i q \cdot X} (i)   
\, \prod_{j=1}^{m}  i a_{M_j} \, \dot{X}^{M_j} \,  e^{i k_j \cdot X} (y_j)}
\\
& = E \; \int_{- \infty}^{\infty} \prod_{i=2}^{m} dy_i \, A_n \,  
\vev{\prod_{j=1}^m  a_{M_j} \, \left[ V^{M_j} (y_j) + q^{M_j} (y_j) \right]}
\end{split}
\end{equation} 
where $A_n$ is given by~\eqref{corr1} and $V^{M_j}$ by~\eqref{bn}. The
expectation values of $q^{M_j}$ are given by the sum over all contraction 
using 
\begin{equation} 
\vev{q^{M_j} (y_j) q^{M_k} (y_k)} = 2 \apr G^{M_j M_k} \frac{1}{(y_j -y_k)^2}.
\end{equation}
In~\eqref{cn} there is a term which is proportional to
}
\begin{multline} \label{dn}
\int_{- \infty}^{\infty} dy \, A_n \, 
\prod_{b=1}^m \frac{i}{\pi} M (k_b)  \frac{1}{1+y_b^2} \\
=  (2 \pi)^d  \delta(\sum_a k_a + q_{\parallel}) \,
\int_{- \infty}^{\infty} dy \, 
\biggl\{ \prod_{a< b} \left[
|\sin (\pi \tau_{ab}) |^{2 \apr k_a \cdot k_b} \exp [\frac{i}{2} 
(k_a \times k_b) (2 \tau_{ab} - \epsilon (\tau_{ab}))] \right] \biggr. \\
\biggl. \times \prod_{b=1}^m i M (k_b)  \frac{1}{\pi (1+y_b^2)}  
\biggr\}
\end{multline}
where $y_a = - \cot \pi \tau_a$ and in the first line 
$A_n$ is given by~\eqref{corr1}. 
Since the integrand~\eqref{dn} is regular for all values 
of $y_a$ we can take the limit 
$\apr k_a \cdot k_b  \ra 0$ {\it before} integrating over the vertex 
operator positions%
\footnote{
Although we are interested
in the $\alpha'  k_a \cdot k_b  \rightarrow 0$ limit, we cannot 
generally take this limit until after integrating, because the 
integrals are typically
defined only by analytic continuation from outside the kinematic
region of interest. We can take  $\apr k_a \cdot k_b  \ra 0$ 
inside the integrals only when the integrand is regular 
throughout the limit.}.
Using $\frac{dy}{d \tau} = \pi (1 + y^2)$ to change the integration 
variables to $\tau_a$ we find 
\begin{equation}
(2 \pi)^{p+1}   \delta(\sum_a k_a + q_{\parallel}) \,
\prod_{b=1}^m i M (k_b) \; \left(\prod_{a=2}^{m} \int_{0}^{1} d\tau_a \right)
\; \prod_{a<b=1}^m \exp [\frac{i}{2} 
(k_a \times k_b) (2 \tau_{ab} - \epsilon (\tau_{ab}))] \ . 
\end{equation}
This is precisely the $m$th-order term in the expansion of a Wilson 
line~\eqref{wils}--\eqref{defjn} with no operator insertion%
\footnote{Recall that the $SL(2,\mathbb{R})$-invariance was used to
fix $y_1=0$ or $\tau_1=\half$.}. 
The factor of $\frac{1}{m!}$ also arises correctly from the 
symmetry factor when translating the amplitude into the effective action.

As a more complicated  example let us look at the term
\begin{equation}
E(q) \;  \int d^{p+1} x  \left[W(x,C) F_{\mu \nu} F^{\mu \nu}\right]
\ast e^{i q_{\rho} x^{\rho}}  
\end{equation}
in~\eqref{effa}, which has an  expansion, using~\eqref{wils}--\eqref{defjn},
of 
\begin{equation} \label{expf}
 \sum_{m=0}^{\infty} 
\frac{1}{m!}  \left( \prod_{a=1}^{m+2} 
\int \! \frac{d^{p+1} k_a}{(2 \pi)^{p+1}} \right) \,
F_{\mu \nu} (k_1) F^{\mu \nu} (k_2) \; \left(\prod_{b=3}^{m+2}  i M(k_{b})
\right)  \,  J_{m+2} (k_1, \cdots, k_{m+2})
\end{equation}
where we have omitted the factor $(2 \pi)^{p+1} \delta
(q + \sum_{a=1}^{m+2} k_a)$ so as not to make the formula too long.
To reproduce the $m$th order term in~\eqref{expf}  we shall look at the 
following amplitude
\begin{equation} \label{fn}
 \int_{- \infty}^{\infty} d y_2 \, \prod_{a=3}^{m+2} dy_a \, 
\vev{V^{-1,-1}_{NSNS} (i) \;\; V^0_O (0;k_1,a_1) \,  V^0_O (y_2;k_2,a_2)
\prod_{a=3}^{m}  V^0_O (y_a;k_a,a_a)}
\end{equation} 
where we have separated two open string vertex operators with the 
rest. For vertex operators $a=3, \cdots, m+2$ we again take only the factors  
$iM_{a}$ while the first two are to be contracted with the closed 
string vertex operator in a way that gives rise to the factor  
$F_{\mu \nu} (k_1) F^{\mu \nu} (k_2) = 2 (k_1 \cdot a_2) (k_2 \cdot a_1)
- 2 (k_1 \cdot k_2) (a_1 \cdot a_2)$.
With this in mind it is easy to extract terms of the 
relevant kinematic structure from the amplitude~\eqref{fn}
\begin{multline}
E (q) \int_{- \infty}^{\infty} \left(d y_2  \, \prod_{i=a}^{m+2} dy_a \right)
\, A_{m+2}\,
\left[ 2\apr  (a_1 \cdot k_2)(a_2 \cdot k_1) \frac{1}{1 + y_2^2} 
+ \frac{1}{y_2^2} (1  + \apr t) 
(a_1 \cdot a_2) \right] \;
\\* \times \prod_{a=3}^{m+2} \frac{iM_a}{\pi (1 + y_a^2)} 
\end{multline}
where $t = - 2 k_1 \cdot k_2$ and 
$A_{m+2}$ is again given by~\eqref{corr1}.
The first term in the square bracket is regular as we take 
$\apr k_a \cdot k_b \ra 0$ and after changing the integration coordinates 
to $\tau_a$ we find, 
\begin{equation} \label{v1}
2 \pi \apr \, E (q) \, (k_1 \cdot a_2) (k_2 \cdot a_1) 
\left(\prod_{j=3}^{m+2} iM_j \right)\; J_{m+2} (k_1, \cdots, k_{m+2})
\end{equation}
The second term has a double pole in $y_2$ as we take $\apr k_1 \cdot k_2$ to 
zero. Thus we can not take the $\apr k_1 \cdot k_2 \ra 0$ na\"{\i}vely 
in the integrand and have to be more careful with the integral over $y_2$. 
Taking all other $\apr k_i \cdot k_j$ to zero except $\apr k_1 \cdot k_2 $ 
and changing the coordinates to $\tau_i$ we get%
\footnote{Recall that  $y_1 = 0$ and so $\tau_1 = \ha$.}
\begin{multline} \label{v2}
E (q)  \, \pi \, (a_1 \cdot a_2) \prod_{a=3}^{m+2} iM_a
(1 + \apr t)
\left(\int_{0}^{1} d \tau_2 \,  \prod_{a=3}^{m+2} d\tau_a \right) \;   
|\cos \pi \tau_2|^{-\apr t -2}
\\* \times
\prod_{a<b=1}^{m+2} \exp [\frac{i}{2} 
(k_a \times k_b) (2 \tau_{ab} - \epsilon (\tau_{ab}))].
\end{multline}
Note an identity
\begin{equation} \label{ibp}
(1 + \apr t) 
|\cos \pi \tau_2|^{-\apr t  - 2} 
= \apr t  |\cos \pi \tau_2 |^{-\apr t   } + 
\frac{1}{\pi^2  \apr t} 
\del_{\tau_2 }^2 |\cos \pi \tau_{2}|^{-\apr t}
\end{equation}
Now both terms on the right hand side are regular in $\tau_2$ 
as we take $\apr  k_1 \cdot k_2 \ra 0$, 
where the second term should be evaluated using integration by parts.   
Since integrating by parts in the second term does not yield anything
to cancel the $1/(k_1 \cdot k_2) $ factor, it contributes a pole term in 
the final amplitude  and is not of interest to us.
Thus in the low energy limit~\eqref{v2} becomes  
\begin{equation} \label{v3}
- 2 \apr \, E \, (a_1 \cdot a_2) (k_1 \cdot k_2) 
\prod_{a=3}^{m+2} iM_a \left( 
\int_{0}^{1} \!  \prod_{a=2}^{m+2} d\tau_a \right) \,   
 \prod_{a<b=1}^{m+2} \exp [\frac{i}{2} 
(k_a \times k_b) (2 \tau_{ab} - \epsilon (\tau_{ab}))] \ .
\end{equation}
Combining~\eqref{v1} and~\eqref{v3} we find precisely~\eqref{expf} including 
the symmetric factor.
We have again suppressed the Chan-Paton factors, but they can again be
easily inserted.  In particular, when $B=0$, only the second term
in~\eqref{getM} survives, which leads to the fully symmetrized trace
over the Chan-Paton factors.  We also note that if at the
beginning we had taken the
large $B$ limit of section~\ref{sec:mt}, then the $F^2$ term we have
just finished considering would not
have appeared.

Similarly we can recover all higher order terms in~\eqref{effa}, by 
specifying a contraction as in the corresponding term 
in~\eqref{preeff} by extracting  factors of  $i M_a$ from other
contractions.

\section{Discussion and Conclusions} \label{sec:conc}

In this paper we have investigated the supergravity couplings of the 
noncommutative D-branes. The general picture that emerges agrees very well 
with the proposal in~\cite{liu} and with the expectations from the Matrix 
Model\cite{dastrivedi}: supergravity fields couple to an open Wilson line 
and additional  operator insertions  have to be
integrated along the Wilson line. 
When $B=0$, this prescription reduces to the 
``symmetrized  trace'' prescription for multiple D-branes interacting 
with background supergravity fields, and the supergravity fields are seen   
to be  functionals of the nonabelian scalar fields on the branes. 

When $B \neq 0$,  the graviton, antisymmetric tensor and the dilaton
all couple to Yang-Mills fields at tadpole, linear and quadratic orders. 
Most of these couplings disappear  
when $B=0$. Then, for example, the dilaton only couples to 
the D3-brane at quadratic level.  Since, from the 
worldsheet point of view turning 
on a constant $B$-field simply corresponds to a relative Lorentz rotation 
between the left and right moving sector, we may consider  the couplings 
at $B \neq 0$  as certain kind of rotation of the original $B = 0$ couplings. 
Indications of this can be seen from the Lorentz index structure 
of~\eqref{e1}--\eqref{e3} (see also equation~\eqref{mixc} below). 
This point of view may be helpful in understanding the 
mixings in the perturbations of supergravity background 
in the noncommutative version of AdS/CFT~\cite{hi,mr}.

Finally we comment on the relation between our results and the 
Born-Infeld action. The effective action~\eqref{effe}--\eqref{cwilson} 
can be obtained from the ordinary  Born-Infeld action by the following 
procedure:
\begin{enumerate}
\item 
Take the Born-Infeld action 
\begin{equation} \label{bi}
S_{BI} = \frac{1}{g_{YM}^2  (2 \pi \apr)^2} \int d^4 x 
\sqrt{\det (G + 2 \pi \apr F)}
\end{equation} 
and perturb the open string metric by $G \ra G + \kappa_{10} \epsilon$, 
where $\epsilon$ does not have to be symmetric. 

\item 
Expand the Born-Infeld action~\eqref{bi} to linear level in $\epsilon$
and quadratic level in $F$ to find that 
\begin{equation} \label{exbi}
\begin{split}
S_{BI} & = \frac{\kappa_{10}}{g_{YM}^2 (2 \pi \apr)^2} 
\int d^4 x \sqrt{\det G}
\, \biggl\{ \frac{1}{2} G^{\mu \nu} \epsilon_{\mu \nu}  \;  \left(
1 + \frac{1}{4} 
F_{\mu \nu} F^{\mu \nu} + 
\half 
D_{\mu} \phi_i D^{\mu} \phi^i \right) \\
& + \ha (G^{\mu \lam} \epsilon_{\lam \tau} G^{\tau \nu}) \;  
\left(- F_{\mu \nu} + F_{\mu\rho} F^{\rho}_{\, \nu} - D_{\mu} \phi^i  D_{\nu} 
\phi_i \right) \biggr\} \ .
\end{split}
\end{equation}

\item
Fourier transform $\epsilon$
\begin{equation} \label{efr}
\epsilon_{\mu \nu} (x) = \int \frac{d^{10} q}{(2 \pi)^{10}} \; 
\epsilon_{\mu \nu} (q)  \exp \left[ i q_{\mu} \cdot X^{\mu} + i 
q_{\perp} \cdot X \right] \ .
\end{equation}

\item 
Turn the exponential factor in~\eqref{efr} into an open Wilson 
line~\eqref{cwilson}, and  at the same time 
integrate each operator in~\eqref{exbi} along the Wilson line with
respect to the $\ast$-multiplication and path ordering. 

\end{enumerate}

Following the above steps we precisely recover~\eqref{effe}--\eqref{cwilson}
with the identification
\begin{equation} \label{mixc}
\epsilon_{\lam \rho} G^{\lam \mu}  G^{\rho \nu} = 
E^{\mu \nu} = e_{\lam \rho} \left(\frac{1}{g + B} \right)^{\lam \mu}
\left(\frac{1}{g - B} \right)^{\rho \nu}
\end{equation}

It would be interesting to check whether  
higher order terms in~\eqref{bi} also reproduce the 
string theory amplitudes  following the same steps. 
There are some indications that this might be true:
\begin{itemize}
\item 
Note that the general disk amplitude~\eqref{namp}  can be written as
\begin{equation}
{\mathcal A}_m = \hat{{\mathcal A}}_m + \tilde{{\mathcal A}}_m
\end{equation}
with 
\begin{equation} \label{gs}
\begin{split}
\hat{{\mathcal A}}_m & = E  \int dy \; Q_m (\xi_i, k_i, y_i;G, \theta) \\
\tilde{{\mathcal A}}_m & = E^{M N} \int dy \; 
          T_{M N}^{(m)} (\xi_i, k_i, y_i;G, \theta)
\end{split}
\end{equation}
where 
\begin{equation} 
E =  e_{M N } \left(\frac{g-B}{g+B} D  \right)^{M N}, \qquad
E^{M N} = e_{P  Q} \left(\frac{1}{g+B}\right)^{P M}
\left(\frac{1}{g- B } D \right)^{Q N}.
\end{equation}
$\Hat{\mathcal A}_m$ comes from the self-contraction of $\psi$ and
$\tilde{\psi}$ and $\Tilde{\mathcal A}_m$ comes from the other terms.
It is important to note that except for the factor $e^{i q \cdot X}$ in the 
closed string vertex $Q_m$ and $T_{MN}^{(m)}$ involve only the contractions 
of the open string vertex operators and  are naturally expressed in
terms of the open string metric $G$ and the noncommutative 
parameter $\theta$. The general dependence on the closed string perturbations
in~\eqref{gs} is consistent with that in~\eqref{exbi} with the 
identification~\eqref{mixc}.

\item 
The additional structures~\eqref{get*n} and~\eqref{getM} 
in the worldsheet correlation functions due to the presence of the $B$-field
is consistent with the appearance of the Wilson line and the integration
procedure as we outlined in Sec.~\ref{sec:wl}.

\end{itemize}

However we note  from the examples in Sec.~\ref{sec:wl} that  the 
emergence of the Wilson line and the integration prescription (and 
similarly the symmetrized trace prescription) depend crucially on 
the analytic properties of the worldsheet 
correlation functions in the low energy limit. 
It would be interesting to investigate them 
systematically. It should be helpful in understanding 
Matrix Theory and also  D-branes in curved space.

\acknowledgments


This work was supported by DOE grant
\hbox{\#DE-FG02-96ER40559}.

\end{document}